\documentclass{moriond}

\def\Journal#1#2#3#4{{#1} {\bf #2}, #3 (#4)}

\def\PRD{{\em Phys. Rev.} D}

\def\be{\begin{equation}}
\def\ee{\end{equation}}
\def\bea{\begin{eqnarray}}
\def\eea{\end{eqnarray}}

\def\sPlot{\mbox{\em sPlot}\xspace}

\usepackage{gensymb} 
\usepackage{amsmath} 
\usepackage{booktabs} 
\usepackage{tablefootnote} 
\usepackage{setspace} 

\usepackage{lineno}  

\usepackage{graphicx}  

\usepackage[para,online,flushleft]{threeparttable}

\newcommand{\gRobustCentral}   { \ensuremath{74.0}\xspace}
\newcommand{\gRobustOnesig}    { \ensuremath{[68.2,79.0]}\xspace}
\newcommand{\gRobustTwosig}    { \ensuremath{[61.6,83.7]}\xspace}

\newcommand{\rbRobustCentral}   { \ensuremath{0.0989}\xspace}
\newcommand{\rbRobustOnesig}    { \ensuremath{[0.0939,0.1040]}\xspace}
\newcommand{\rbRobustTwosig}    { \ensuremath{[0.0891,0.1087]}\xspace}

\newcommand{\dbRobustCentral}   { \ensuremath{131.2}\xspace}
\newcommand{\dbRobustOnesig}    { \ensuremath{[125.3,136.3]}\xspace}
\newcommand{\dbRobustTwosig}    { \ensuremath{[118.3,140.9]}\xspace}

\newcommand{\rbDstKpRobustCentral}   { \ensuremath{0.191}\xspace}
\newcommand{\rbDstKpRobustOnesig}    { \ensuremath{[0.153,0.236]}\xspace}
\newcommand{\rbDstKpRobustTwosig}    { \ensuremath{[0.121,0.287]}\xspace}

\newcommand{\dbDstKpRobustCentral}   { \ensuremath{331.6}\xspace}
\newcommand{\dbDstKpRobustOnesig}    { \ensuremath{[321.4,339.8]}\xspace}
\newcommand{\dbDstKpRobustTwosig}    { \ensuremath{[309,346]}\xspace}

\newcommand{\rbDKstpRobustCentral}   { \ensuremath{0.092}\xspace}
\newcommand{\rbDKstpRobustOnesig}    { \ensuremath{[0.059,0.110]}\xspace}
\newcommand{\rbDKstpRobustTwosig}    { \ensuremath{[0.034,0.126]}\xspace}

\newcommand{\dbDKstpRobustCentral}   { \ensuremath{40}\xspace}
\newcommand{\dbDKstpRobustOnesig}    { \ensuremath{[20,132]}\xspace}
\newcommand{\dbDKstpRobustTwosig}    { \ensuremath{[5,155]}\xspace}

\newcommand{\rbDKstzRobustCentral}   { \ensuremath{0.221}\xspace}
\newcommand{\rbDKstzRobustOnesig}    { \ensuremath{[0.174,0.265]}\xspace}
\newcommand{\rbDKstzRobustTwosig}    { \ensuremath{[0.123,0.309]}\xspace}

\newcommand{\dbDKstzRobustCentral}   { \ensuremath{187}\xspace}
\newcommand{\dbDKstzRobustOnesig}    { \ensuremath{[167,210]}\xspace}
\newcommand{\dbDKstzRobustTwosig}    { \ensuremath{[148,239]}\xspace}

\newcommand{\rbDsKRobustCentral}   { \ensuremath{0.301}\xspace}
\newcommand{\rbDsKRobustOnesig}    { \ensuremath{[0.215,0.391]}\xspace}
\newcommand{\rbDsKRobustTwosig}    { \ensuremath{[0.14,0.49]}\xspace}

\newcommand{\dbDsKRobustCentral}   { \ensuremath{355}\xspace}
\newcommand{\dbDsKRobustOnesig}    { \ensuremath{[339,372]}\xspace}
\newcommand{\dbDsKRobustTwosig}    { \ensuremath{[321,390]}\xspace}

\newcommand{\rbDKppRobustCentral}   { \ensuremath{0.081}\xspace}
\newcommand{\rbDKppRobustOnesig}    { \ensuremath{[0.054,0.106]}\xspace}
\newcommand{\rbDKppRobustTwosig}    { \ensuremath{[0.000,0.125]}\xspace}

\newcommand{\dbDKppRobustCentral}   { \ensuremath{351.4}\xspace}
\newcommand{\dbDKppRobustOnesig}    { \ensuremath{[314.0,359.8]}\xspace}
\newcommand{\dbDKppRobustTwosig}    { \ensuremath{[180,360]}\xspace}

\newcommand{\dbDPiRobustCentral}   { \ensuremath{17}\xspace}
\newcommand{\dbDPiRobustOnesig}    { \ensuremath{[0,46]}\xspace}
\newcommand{\dbDPiRobustTwosig}    { \ensuremath{[0,76]}\xspace}

\newcommand{\g}{\texorpdfstring{\ensuremath{\gamma}}{gamma}\xspace}

\newcommand{\rb}  {\texorpdfstring{\ensuremath{r_B^{DK}}}{rBDK}\xspace}

\newcommand{\db}  {\ensuremath{\delta_B^{DK}}\xspace}

\newcommand{\rbDkpp}{\ensuremath{r_B^{DK\pi\pi}}\xspace}

\newcommand{\dbDkpp}{\ensuremath{\delta_B^{DK\pi\pi}}\xspace}

\newcommand{\rbDstKp}{\ensuremath{r_B^{D^{*}K^{+}}}\xspace}
\newcommand{\dbDstKp}{\ensuremath{\delta_B^{D^{*}K^{+}}}\xspace}

\newcommand{\rbDKstp}{\ensuremath{r_B^{DK^{*+}}}\xspace}
\newcommand{\dbDKstp}{\ensuremath{\delta_B^{DK^{*+}}}\xspace}

\newcommand{\rbDKstz}{\ensuremath{r_B^{DK^{*0}}}\xspace}

\newcommand{\dbDKstz}{\ensuremath{\delta_B^{DK^{*0}}}\xspace}

\newcommand{\ddpi}{\texorpdfstring{\ensuremath{\delta_B^{D^\mp\pi^\pm}}\xspace}{delta_Dpi}}

\newcommand{\rdsk}{\texorpdfstring{\ensuremath{r_B^{D_s^\mp K^\pm}}}{rDsK}\xspace}

\newcommand{\ddsk}{\texorpdfstring{\ensuremath{\delta_B^{D_s^\mp K^\pm}}}{dDsK}\xspace}

\newcommand{\hp} {\texorpdfstring{\ensuremath{h^{+}}}{h+}\xspace}
\newcommand{\hm} {\texorpdfstring{\ensuremath{h^{-}}}{h-}\xspace}

\newcommand{\BuDK}    {\texorpdfstring{\ensuremath{\Bu\to D \Kp}}		{B+ -> DK+}\xspace}
\newcommand{\BuDstK}  {\texorpdfstring{\ensuremath{\Bu\to D^* \Kp}}		{B+ -> DstK+}\xspace}
\newcommand{\BuDKst}  {\texorpdfstring{\ensuremath{\Bu\to D K^{*+}}}		{B+ -> DstKst+}\xspace}

\newcommand{\BuDKpipi}{\texorpdfstring{\ensuremath{\Bu\to D \Kp\pip\pim}}		{B+ -> DK+pi+pi-}\xspace}

\newcommand{\BdDpipm}   {\texorpdfstring{\decay{\Bz}{\Dmp \pipm}}		{}}

\newcommand{\BdDKpi}  {\texorpdfstring{\decay{\Bd}{\D\Kp\pim}} {}}
\newcommand{\BdDzKstz}{\texorpdfstring{\ensuremath{\Bd\to D K^{*0}}}		{B0 -> DK*0}\xspace}

\newcommand{\BsDsK}	{\texorpdfstring{\ensuremath{\Bs\to D_s^\mp K^\pm}}	{Bs -> DsK}\xspace}

\newcommand{\Dshhh}    {\decay{\Ds}{\hp\hm\pip}}

\newcommand{\DKpipi}   {\decay{\Dp}{\Kp\pim\pip}}

\newcommand{\DKpi}     {\texorpdfstring{\ensuremath{D\to K^+\pi^-}}{D -> Kpi}\xspace}
\newcommand{\DKSKpi}   {\texorpdfstring{\ensuremath{D\to \KS K^+\pi^-}}{D -> KSKpi}\xspace}

\newcommand{\Dhh}      {\texorpdfstring{\ensuremath{D\to h^+h^-}}{D -> hh}\xspace}

\newcommand{\Dhpipipi} {\texorpdfstring{\ensuremath{D\to h^+\pi^-\pi^+\pi^-}}{D -> hpipipi}\xspace}

\newcommand{\Dhhpiz}   {\texorpdfstring{\ensuremath{D\to h^+h^-\piz}}{D -> hhpi0}\xspace}

\newcommand{\DKSpipi}  {\texorpdfstring{\ensuremath{D\to\KS\pi^+\pi^-}}{D -> KSpipi}\xspace}

\newcommand{\DKShh}    {\texorpdfstring{\ensuremath{D\to\KS h^+h^-}}{D -> KShh}\xspace}

\usepackage{ifthen} 
\newboolean{uprightparticles}
\setboolean{uprightparticles}{false} 
\usepackage{xspace} 
\usepackage{upgreek}

\def\MagUp {\mbox{\em Mag\kern -0.05em Up}\xspace}

\ifthenelse{\boolean{uprightparticles}}%
{
 
 \def\Pgamma      {\ensuremath{\upgamma}\xspace}

 \def\Ppi         {\ensuremath{\uppi}\xspace}

 \def\PDelta      {\ensuremath{\Delta}\xspace}                 
 \def\PXi      {\ensuremath{\Xi}\xspace}                 
 \def\PLambda      {\ensuremath{\Lambda}\xspace}                 
 \def\PSigma      {\ensuremath{\Sigma}\xspace}                 
 \def\POmega      {\ensuremath{\Omega}\xspace}                 
 \def\PUpsilon      {\ensuremath{\Upsilon}\xspace}                 
 
 \def\PB      {\ensuremath{\mathrm{B}}\xspace}                 
                  
 \def\PD      {\ensuremath{\mathrm{D}}\xspace}

 \def\PK      {\ensuremath{\mathrm{K}}\xspace}

 \def\Pi      {\ensuremath{\mathrm{i}}\xspace}

 \def\Ps      {\ensuremath{\mathrm{s}}\xspace}

}
{
 
 \def\Pgamma      {\ensuremath{\gamma}\xspace}

 \def\Ppi         {\ensuremath{\pi}\xspace}

 \mathchardef\PDelta="7101
 \mathchardef\PXi="7104
 \mathchardef\PLambda="7103
 \mathchardef\PSigma="7106
 \mathchardef\POmega="710A
 \mathchardef\PUpsilon="7107
                  
 \def\PB      {\ensuremath{B}\xspace}                 
                  
 \def\PD      {\ensuremath{D}\xspace}

 \def\PK      {\ensuremath{K}\xspace}

 \def\Pi      {\ensuremath{i}\xspace}

 \def\Ps      {\ensuremath{s}\xspace}

}

\makeatletter
\ifcase \@ptsize \relax
  \newcommand{\miniscule}{\@setfontsize\miniscule{4}{5}}
\or
  \newcommand{\miniscule}{\@setfontsize\miniscule{5}{6}}
\or
  \newcommand{\miniscule}{\@setfontsize\miniscule{5}{6}}
\fi
\makeatother

\DeclareRobustCommand{\optbar}[1]{\shortstack{{\miniscule (\rule[.5ex]{1.25em}{.18mm})}
  \\ [-.7ex] $#1$}}

\def\g      {{\ensuremath{\Pgamma}}\xspace}

\def\squark    {{\ensuremath{\Ps}}\xspace}

\def\pion   {{\ensuremath{\Ppi}}\xspace}
\def\piz    {{\ensuremath{\pion^0}}\xspace}

\def\pip    {{\ensuremath{\pion^+}}\xspace}
\def\pim    {{\ensuremath{\pion^-}}\xspace}
\def\pipm   {{\ensuremath{\pion^\pm}}\xspace}

\def\kaon    {{\ensuremath{\PK}}\xspace}
  \def\Kbar    {{\kern 0.2em\overline{\kern -0.2em \PK}{}}\xspace}

\def\KorKbar    {\kern 0.18em\optbar{\kern -0.18em K}{}\xspace}

\def\Kp      {{\ensuremath{\kaon^+}}\xspace}

\def\KS      {{\ensuremath{\kaon^0_{\mathrm{ \scriptscriptstyle S}}}}\xspace}

  \def\Dbar    {{\kern 0.2em\overline{\kern -0.2em \PD}{}}\xspace}
\def\D       {{\ensuremath{\PD}}\xspace}

\def\DorDbar    {\kern 0.18em\optbar{\kern -0.18em D}{}\xspace}

\def\Dp      {{\ensuremath{\D^+}}\xspace}

\def\Dmp     {{\ensuremath{\D^\mp}}\xspace}

\def\Ds      {{\ensuremath{\D^+_\squark}}\xspace}

\def\B       {{\ensuremath{\PB}}\xspace}
\def\Bbar    {{\ensuremath{\kern 0.18em\overline{\kern -0.18em \PB}{}}}\xspace}
\def\Bb      {{\ensuremath{\Bbar}}\xspace}
\def\BorBbar    {\kern 0.18em\optbar{\kern -0.18em B}{}\xspace}
\def\Bz      {{\ensuremath{\B^0}}\xspace}

\def\Bu      {{\ensuremath{\B^+}}\xspace}

\def\Bd      {{\ensuremath{\B^0}}\xspace}
\def\Bs      {{\ensuremath{\B^0_\squark}}\xspace}

  \def\Y#1S{\ensuremath{\PUpsilon{(#1S)}}\xspace}

\def\Lbar        {{\ensuremath{\kern 0.1em\overline{\kern -0.1em\PLambda}}}\xspace}
\def\LorLbar    {\kern 0.18em\optbar{\kern -0.18em \PLambda}{}\xspace}

\newcommand{\decay}[2]{\ensuremath{#1\!\to #2}\xspace}         

\def\to                 {\ensuremath{\rightarrow}\xspace}

\def\AT#1     {\ensuremath{A_{\mathrm{T}}^{#1}}\xspace}           

\def\C#1      {\ensuremath{\mathcal{C}_{#1}}\xspace}                       
\def\Cp#1     {\ensuremath{\mathcal{C}_{#1}^{'}}\xspace}                    
\def\Ceff#1   {\ensuremath{\mathcal{C}_{#1}^{\mathrm{(eff)}}}\xspace}        
\def\Cpeff#1  {\ensuremath{\mathcal{C}_{#1}^{'\mathrm{(eff)}}}\xspace}       
\def\Ope#1    {\ensuremath{\mathcal{O}_{#1}}\xspace}                       
\def\Opep#1   {\ensuremath{\mathcal{O}_{#1}^{'}}\xspace}                    

\newcommand{\tev}{\ensuremath{\mathrm{\,Te\kern -0.1em V}}\xspace}
\newcommand{\gev}{\ensuremath{\mathrm{\,Ge\kern -0.1em V}}\xspace}
\newcommand{\mev}{\ensuremath{\mathrm{\,Me\kern -0.1em V}}\xspace}
\newcommand{\kev}{\ensuremath{\mathrm{\,ke\kern -0.1em V}}\xspace}
\newcommand{\ev}{\ensuremath{\mathrm{\,e\kern -0.1em V}}\xspace}
\newcommand{\gevc}{\ensuremath{{\mathrm{\,Ge\kern -0.1em V\!/}c}}\xspace}
\newcommand{\mevc}{\ensuremath{{\mathrm{\,Me\kern -0.1em V\!/}c}}\xspace}
\newcommand{\gevcc}{\ensuremath{{\mathrm{\,Ge\kern -0.1em V\!/}c^2}}\xspace}
\newcommand{\gevgevcccc}{\ensuremath{{\mathrm{\,Ge\kern -0.1em V^2\!/}c^4}}\xspace}
\newcommand{\mevcc}{\ensuremath{{\mathrm{\,Me\kern -0.1em V\!/}c^2}}\xspace}

\def\invfb   {\ensuremath{\mbox{\,fb}^{-1}}\xspace}

\def\gsim{{~\raise.15em\hbox{$>$}\kern-.85em
          \lower.35em\hbox{$\sim$}~}\xspace}
\def\lsim{{~\raise.15em\hbox{$<$}\kern-.85em
          \lower.35em\hbox{$\sim$}~}\xspace}

\def\sPlot{\mbox{\em sPlot}\xspace}

\def\tell1  {TELL1\xspace}
\def\ukl1   {UKL1\xspace}

\newcommand{\Photo}{\includegraphics[height=35mm]{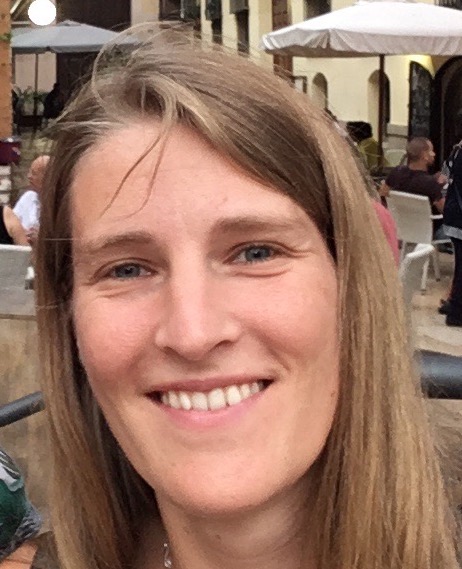}}

\begin{document}
\vspace*{4cm}
\title{Time-integrated $C\!P$-violation in beauty at LHCb}

\author{ Emilie Bertholet on behalf of the LHCb collaboration }

\address{LPNHE, Sorbonne Universit\'e, Paris Diderot Sorbonne Paris Cit\'e, CNRS/IN2P3,\\ Paris, France}

\maketitle\abstracts{
Precision measurements of time-integrated $C\!P$ violation in beauty decays permit a better understanding of the different mechanisms underlying $C\!P$ violation. They allow to better constrain the Standard Model and probe for new physics. A selection of recent LHCb results that highlight different aspects of $C\!P$ violation in $b$-hadron decays are presented here.  }

\section{Update of the LHCb combination of the CKM angle $\gamma$}

The CKM phase $\gamma$ can be measured either in tree dominated decays or in processes that contain a significant contribution from loop diagrams. The latter are potentially sensitive to New Physics (NP) while no significant NP effects are expected in the former. Thus, the tree-level measurements constitute a benchmark for the Standard Model (SM). The comparison of the results obtained via these two approaches give valuable input to constrain the SM and set limits on NP.

This analysis~\cite{LHCb-CONF-2018-002} combines several tree-level LHCb measurements of $\gamma$ in $B \to D h$ decay modes, where $h$ represents a hadron. The different methods to extract $\gamma$ exploit the interference between $b\to c$ (favoured) and $b\to u$ (suppressed) transitions. The ratio of the corresponding amplitudes is related to $\gamma$ by
\be
\label{gamma_tree}
\frac{A_{b\to u}}{A_{b\to c}} = r_{B}^{Dh}e^{\delta_{B}^{Dh} \pm \gamma},
\ee
where $r_{B}^{Dh}$ is the ratio of magnitudes, $\delta_{B}^{Dh}$ the strong phase difference between $A_{b\to u}$ and $A_{b\to c}$, and the +(-) sign is associated with the decay of a meson containing a $\overline{b}$ ($b$) quark. Different methods are employed depending on the decay channel of the $D$-meson.
The theoretical uncertainties on such tree-level determination are very small, thus the uncertainties on $\gamma$ depend mainly on the experimental precision. As the $B \to D h$ decay modes have low branching ratios, the best precision on $\gamma$ is obtained by combining results from many decay modes. A list of all the modes used by LHCb in this combination along with the status of the analyses since the last combination is given in Table~\ref{tab:inputs}.

\begin{table}[h!]
\caption{List of the LHCb measurements used in the combination. }

  \begin{center}
    \begin{threeparttable}
      \renewcommand{\arraystretch}{0.9}
      \begin{tabular}{l l l l p{3.2cm}}
        \hline
        $\B$ decay  & $\D$ decay & Method  & Dataset\tnote{\hyperlink{tabnote}{$^\dagger$}} & Status since last combination~\cite{LHCb-CONF-2017-004}  \\
        \hline
        \BuDK     & \Dhh      & GLW       & Run 1 \& 2  & Minor update       \\
        \BuDK     & \Dhh      & ADS       & Run 1       & As before          \\
        \BuDK     & \Dhpipipi & GLW/ADS   & Run 1       & As before          \\
        \BuDK     & \Dhhpiz   & GLW/ADS   & Run 1       & As before          \\
        \BuDK     & \DKShh    & GGSZ      & Run 1       & As before          \\
        \BuDK     & \DKShh    & GGSZ      & Run 2       & New                \\
        \BuDK     & \DKSKpi   & GLS       & Run 1       & As before          \\
        \BuDstK   & \Dhh      & GLW       & Run 1 \& 2  & Minor update       \\
        \BuDKst   & \Dhh      & GLW/ADS   & Run 1 \& 2  & Updated results    \\
        \BuDKst   & \Dhpipipi & GLW/ADS   & Run 1 \& 2  & New                \\
        \BuDKpipi & \Dhh      & GLW/ADS   & Run 1       & As before          \\
        \BdDzKstz & \DKpi     & ADS       & Run 1       & As before          \\
        \BdDKpi   & \Dhh      & GLW-Dalitz   & Run 1       & As before          \\
        \BdDzKstz & \DKSpipi  & GGSZ      & Run 1       & As before          \\
        \BsDsK    & \Dshhh    & TD        & Run 1       & Updated results    \\
        \BdDpipm  & \DKpipi   & TD        & Run 1       & New                \\
        \hline
      \end{tabular}
      \begin{tablenotes}
        \hypertarget{tabnote}{\footnotesize $^\dagger$ Run 1 corresponds to an integrated luminosity of 3\invfb taken at centre-of-mass energies of 7 and 8\tev. Run 2 refers to the data collected in 2015 and 2016, which corresponds to an integrated luminosity of 2\invfb taken at a centre-of-mass energy of 13\tev.}
      \end{tablenotes}
    \end{threeparttable}
  \end{center}
 
  \label{tab:inputs}
\end{table}

The result is obtained using a frequentist approach, following the strategy of the previous combination~\cite{LHCb-CONF-2017-004}, with auxiliary inputs coming form HFLAV, CLEO and LHCb. The likelihood function is built from the product of probability density functions of 98 experimental observables, and 40 parameters are left free in the fit. The hadronic parameters $r_{B}^{Dh}$ and $\delta_{B}^{Dh}$, defined in
Eq.~\eqref{gamma_tree}, are also extracted along with $\gamma$. Table~\ref{tab:resultrobust} also gives a summary of the central values and confidence levels for the parameters of interest.

The combination results in $\gamma = (74.0^{+5.0}_{-5.8})^{\degree}$, including both statistical and systematic uncertainties. This result supersedes the previous LHCb combination and consists in the most precise determination of $\gamma$ from a single experiment to date. 

\begin{table}[!h]
\centering
\caption{Confidence intervals and central values for the parameters of interest.}
\label{tab:resultrobust} 
\begin{tabular}{lccc}
\hline
Quantity & Value & 68.3\% CL & 95.5\% CL \\
\hline
\g [$^\circ$]      & \gRobustCentral & \gRobustOnesig & \gRobustTwosig \\
\rb                & \rbRobustCentral & \rbRobustOnesig & \rbRobustTwosig \\
\db [$^\circ$]     & \dbRobustCentral & \dbRobustOnesig & \dbRobustTwosig \\
\rbDstKp           & \rbDstKpRobustCentral & \rbDstKpRobustOnesig & \rbDstKpRobustTwosig \\
\dbDstKp [$^\circ$]& \dbDstKpRobustCentral & \dbDstKpRobustOnesig & \dbDstKpRobustTwosig \\
\rbDKstp           & \rbDKstpRobustCentral & \rbDKstpRobustOnesig & \rbDKstpRobustTwosig \\
\dbDKstp [$^\circ$]& \dbDKstpRobustCentral & \dbDKstpRobustOnesig & \dbDKstpRobustTwosig \\
\rbDKstz           & \rbDKstzRobustCentral & \rbDKstzRobustOnesig & \rbDKstzRobustTwosig \\
\dbDKstz [$^\circ$]& \dbDKstzRobustCentral & \dbDKstzRobustOnesig & \dbDKstzRobustTwosig \\
\rbDkpp            & \rbDKppRobustCentral & \rbDKppRobustOnesig & \rbDKppRobustTwosig \\
\dbDkpp [$^\circ$] & \dbDKppRobustCentral & \dbDKppRobustOnesig & \dbDKppRobustTwosig \\
\rdsk              & \rbDsKRobustCentral & \rbDsKRobustOnesig & \rbDsKRobustTwosig \\
\ddsk [$^\circ$]   & \dbDsKRobustCentral & \dbDsKRobustOnesig & \dbDsKRobustTwosig \\
\ddpi [$^\circ$]   & \dbDPiRobustCentral & \dbDPiRobustOnesig & \dbDPiRobustTwosig \\
\hline
\end{tabular}

\end{table}

\section{Amplitude analysis of $B^{\pm} \to \pi^{\pm}K^+K^-$}

Previous LHCb analysis of $B\to hh'h''$ decay modes~\cite{previousLHCb} reported localised $C\!P$ asymmetries in some regions of the Dalitz plane (DP). In particular, significant positive (negative) $C\!P$ asymmetry was seen in the $K^+K^-$ ($\pi^+\pi^-$) invariant mass region below 1.5~GeV/$c^2$. These asymmetries, not clearly related to any resonant component, could be due to long-distance $\pi\pi \leftrightarrow KK$ hadronic rescattering. Better understanding of these effects require Dalitz plot analyses.

A DP amplitude analysis of $B^{\pm} \to \pi^{\pm}K^+K^-$ decays is performed for the first time~\cite{B2piKK}, using 3~fb$^{-1}$ of data collected by the LHCb experiment at centre-of-mass energies of 7~TeV and 8~TeV. This analysis uses the isobar model, which gives a description of the decay amplitude as a function of the DP coordinates ($m^2_{\pi^{\pm} K^{\mp}}, m^2_{K^{\pm}K^{\mp}}$) within a quasi two-body approach:
\be
A(m^2_{\pi^{\pm} K^{\mp}}, m^2_{K^{\pm}K^{\mp}}) = \sum_{j=1}^{\rm{nRes}} c_j F_j(m^2_{\pi^{\pm} K^{\mp}}, m^2_{K^{\pm}K^{\mp}}), 
\ee
where the index $j$ runs over the $\rm{nRes}$ components included in the model, $F_j$ are functions that describe the momentum-dependent part of the strong dynamics and the coefficients $c_j$ are the so-called isobar parameters. These are complex numbers that describe the weak interaction and the momentum-independent part of the strong interaction. The information on $C\!P$ violation is encoded into these coefficients so that the $C\!P$ asymmetry for each contribution can be obtained by
\be 
\mathcal{A}_{C\!P, i} = \frac{|\bar{c}_i|^2 - |c_i|^2}{|\bar{c}_i|^2 + |c_i|^2},
\ee 
where the index $i$ designates one of the isobar components. Other relevant observables are the fit fractions, which are the ratio of the integral of one partial amplitude squared, $|A_i|^2$, over the integral of the total amplitude squared. They relate to the relative rate of an isobar component. The flavour-averaged fit fractions are given by
\be 
FF_i = \frac{\iint \left( |c_iF_i(m^2_{\pi^{\pm} K^{\mp}}, m^2_{K^{\pm}K^{\mp}})|^2 +| \bar{c}_i\bar{F}_i(m^2_{\pi^{\pm} K^{\mp}}, m^2_{K^{\pm}K^{\mp}})|^2 \right)\mathrm{d} m^2_{\pi^{\pm}K^{\mp}} \mathrm{d} m^2_{K^{\pm}K^{\mp}} }{ \iint \left( |A|^2 +| \bar{A}|^2 \right)\mathrm{d} m^2_{\pi^{\pm}K^{\mp}} \mathrm{d} m^2_{K^{\pm}K^{\mp}}  }.
\ee

After a careful selection of the candidates, a fit to the invariant ($\pi KK$) mass is performed to obtain the $B^{\pm}$ signal yields. The amplitude model is then built in two steps. At first, all the known resonances that may contribute to the final state are included. The model is then further refined by adding or removing components following a systematic procedure in order to find the configuration that best describes the data. Seven contributions to the total amplitude are retained in the final result. Among them, a rescattering component~\cite{rescatt} is found to give a good description of the data in the invariant mass window $1.0\; {\rm GeV} < m_{KK} < 1.5\;{\rm GeV}$. The non-resonant contribution is described by a polar form factor~\cite{polar}, which is a phenomenological description of the partonic interaction that produces the final state. Five resonances are also included: $K^{*}(892)^0$, $K^{*0}(1430)$, $\rho(1450)$, $f_2(1270)$ and $\phi(1020)$.

The dominant contribution is found to originate from the non-resonant component, with a fit fraction of 32.3\%. A significant contribution, 16.4\%, from the rescattering component is also observed, along with a very small, non significant, contribution from the $\phi(1020)$, 0.3\%. The rescattering amplitude comes with a very large negative $C\!P$ asymmetry, $-66.4\%$, which seems to indicate that the large localised $C\!P$ asymmetries observed previously can be explained through $\pi \pi \leftrightarrow K K$ rescattering.

\section{Study of $B^0 \to \rho(770)^0K^*(892)^0$}

The study $B^0 \to \rho(770)^0K^*(892)^0$, through a full amplitude analysis of the 4-body $(\pi^+ \pi^-)(K^+ \pi^-)$ final state, is performed for the first time. The analysis uses 3~fb$^{-1}$ of data collected by the LHCb experiment at centre-of-mass energies of 7~TeV and 8~TeV. Three leading-order diagrams contribute to the final state: the tree-level contribution is doubly Cabbibo suppressed so that the dominant contributions comes from gluonic and electoweak (EW) penguins, which have similar sizes. Additionally, the sign of the EW-penguin contribution depends on the helicity eigenstate, which can have an impact on the value of the polarisation fraction. Furthermore, an enhanced direct $C\!P$-violating effect is expected due to the interference with $B^0 \to \omega K^*(892)^0$ decay mode~\cite{omegaK}. Finally, the study of $B\to VV$ modes can help to understand the so-called polarisation puzzle: using na\"ive arguments from the quark helicity conservation and the V-A nature of the weak interaction one expects very large polarisation fractions for $B$ decays into light vector mesons. This turns out to hold for tree dominated decays but not for penguin dominated decays. Recent calculations in perturbative QCD~\cite{pQCD} and QCD factorisation~\cite{QCDF} can accommodate for low longitudinal polarisation fractions in penguin-dominated decays by taking into account a strong-interaction effect. 

\begin{figure}[bthp]
  \begin{center}
  \begin{tabular}{cc}
    \includegraphics[width=7.6cm]{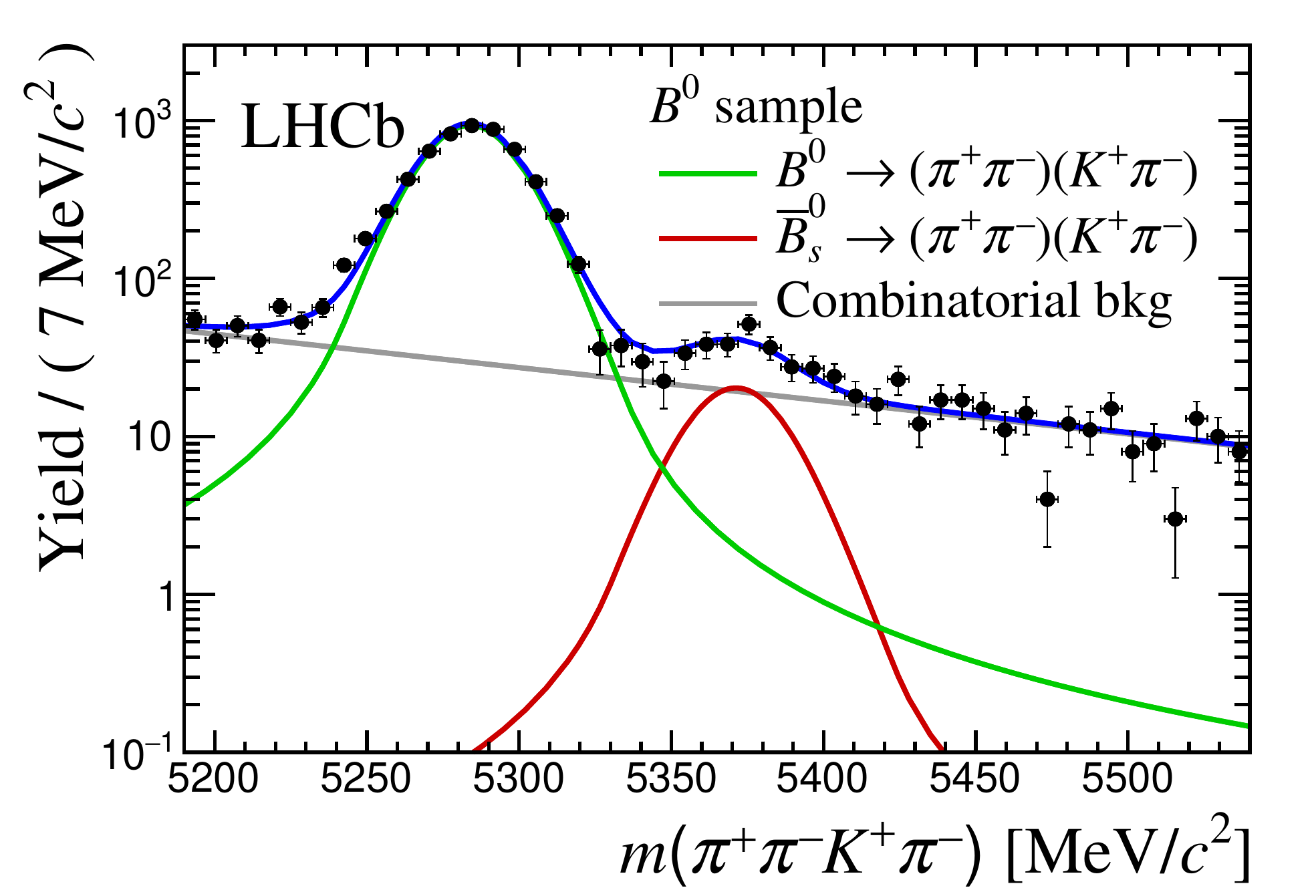} &
    \includegraphics[width=7.6cm]{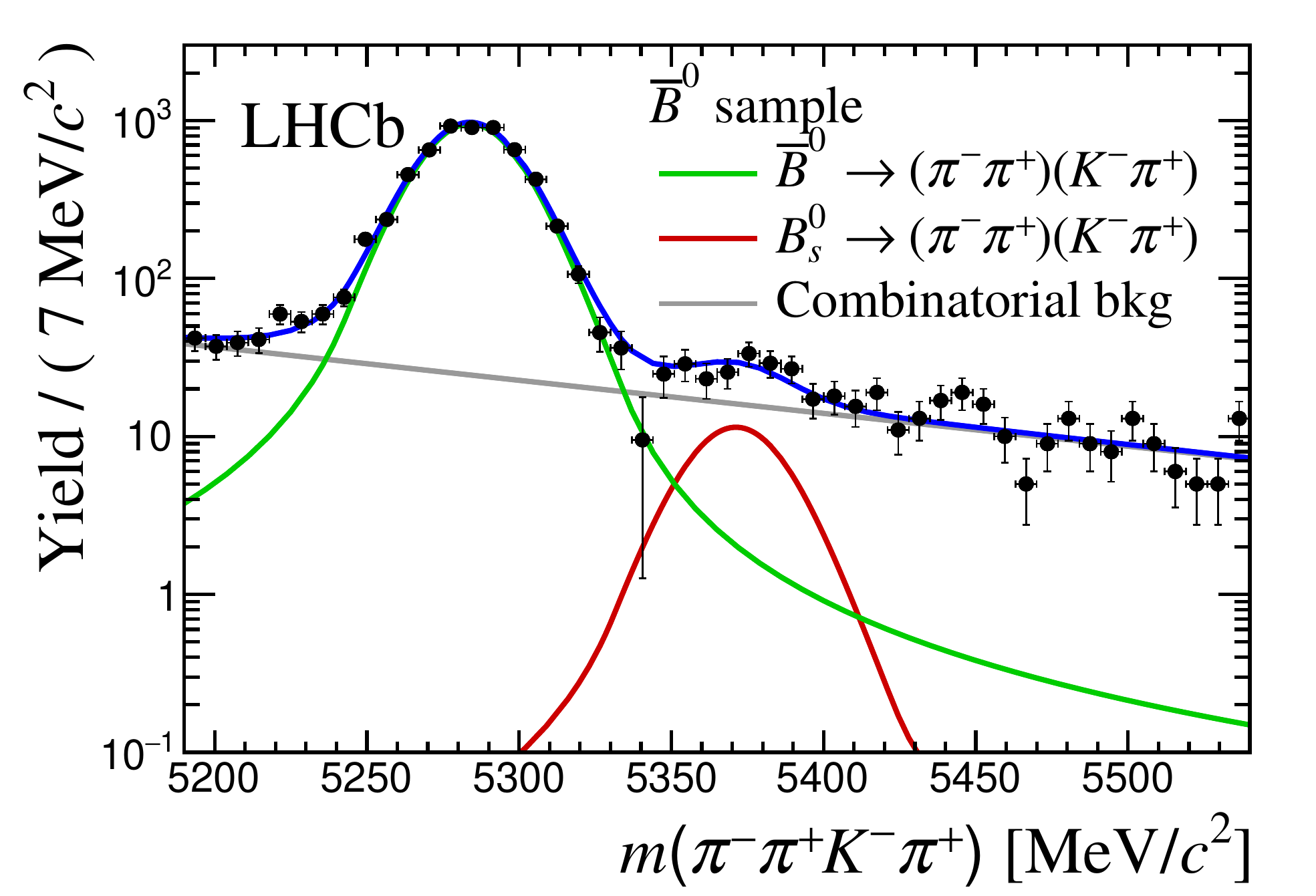} \\
    \small (a) &
    \small (b)
  \end{tabular}
  \end{center}
  \caption{Fit to the invariant-mass distributions of the selected (a) \B and (b) \Bb candidates. }
\label{rockstar_massfit}
\end{figure}

The best candidates are retained by applying trigger requirements and a selection based on topological variables. Cross-feed and combinatorial backgrounds are further reduced by using particle identification and multivariate analysis. The $(\pi^+ \pi^-)$ and $(K^+ \pi^-)$ candidates are selected within invariant mass windows around the masses of the $\rho$ and the $K^*$ resonances: $300\;{\rm MeV}/c^2 < m_{\pi \pi} < 1100\;{\rm MeV}/c^2$ and $750\;{\rm MeV}/c^2 < m_{\pi \pi} < 1200\;{\rm MeV}/c^2$. A fit to the 4-body invariant mass spectrum is then performed (see Fig.\ref{rockstar_massfit}), and the \sPlot~\cite{sPlot} technique is used to obtain background-subtracted samples. These background-subtracted samples are then used to perform a full amplitude analysis.

The final state can be described by using combinations of S-waves (spin 0) and P-waves (spin 1). The contributions to the total amplitude included in the fit are listed in Table~\ref{rockstar_contributions}; they correspond to resonances that are expected in the $(\pi\pi)$ and $(K\pi)$ channels considering the invariant mass regions selected. The amplitude is then built by combining the different contributions together; the final state can thus  be vector-vector (VV), vector-scalar (VS), scalar-vector (SV) or scalar-scalar (SS). In the case of the VV final state, three amplitudes with different polarisations contribute to the decay rate: longitudinal $A_{L}$, parallel $A_{||}$ or transverse $A_{\perp}$. A total of fourteen amplitudes is accounted for and modelled using the isobar model. An angular analysis is needed to study the 4-body final state so that the decay-rate is five-dimensional (two invariant masses and three helicity angles)
\be
\frac{\rm{d}^5 \Gamma}{\rm{d}\cos\theta_{\pi\pi} \rm{d}\cos\theta_{K\pi} \rm{d}\phi \rm{d} m_{\pi\pi} \rm{d}m_{K\pi}} \propto \Phi_{4}(m_{\pi\pi}, m_{K\pi}) \left| \sum_{i}A_iR_i(m_{\pi\pi}, m_{K\pi})g_i(\theta_{\pi\pi}, \theta_{Ki\pi}, \phi) \right|^2,
\ee
where $\Phi_{4}(m_{\pi\pi}, m_{K\pi})$ is the four-body phase-space density, $R_i$ and $g_i$ correspond to the mass and the angular distributions, respectively, and $A_i$ are the decay amplitudes for each component $i$. 

\begin{table}
\label{rockstar_contributions}
\caption{List of the different contributions to the total amplitude.}
\begin{center}
\begin{tabular}{c|cc}
  &
\multicolumn{1}{c}{$(\pi \pi)$} &
\multicolumn{1}{c}{$(K \pi)$} \\
\midrule
Scalar  & $f_0(500)$, $f_0(980)$, $f_0(1300)$ & $K_0^*(1430)^0$+NR \\
Vector & $\omega$, $\rho^0(770)$ & $K^*(892)^0$  \\
\end{tabular}
\end{center}
\end{table}

Polarisation fractions are computed for the the VV final state
\be
f_{\lambda} = \frac{|A_{VV}^{\lambda}|^2}{|A_{VV}^{L}|^2+|A_{VV}^{||}|^2+|A_{VV}^{ \perp}|^2},
\ee
where $\lambda$ represent one of the polarisation configurations. The $C\!P$-averaged fraction, $\tilde{f}^{\lambda} =  (f_{\lambda} + \bar{f}_{\lambda})/2$, and $C\!P$ asymmetries, $\mathcal{A}_{C\!P}^{\lambda} =  (\bar{f}_{\lambda}-f_{\lambda}) / (f_{\lambda} + \bar{f}_{\lambda})$, can be obtained from the polarisation fractions of a mode and its conjugate. $C\!P$-averages and asymmetries are measured for each amplitude included in this analysis. Detailed results can be found in Ref.~\cite{rockstar}. 

 A small longitudinal polarisation fraction and a rather large $C\!P$ asymmetry are measured for the $B^0 \to \rho(770)^0K^*(892)^0$ mode
\be 
\tilde{f}^L_{\rho K^*} = 0.164\pm0.015\pm0.022, ~~~~ \mathcal{A}^L_{\rho K^*} = -0.62\pm0.09\pm0.09.
\ee
These results hint for a relevant contribution from the EW-penguin diagram. The significance of the $C\!P$ asymmetry is about five standard deviations, which consists in the first significant observation of $C\!P$ asymmetry in angular distributions of $B^0 \to VV$ decays. Comparison of these results to the most recent theoretical predictions pQCD~\cite{pQCD} and QCDf~\cite{QCDF} show a good agreement. The longitudinal polarisation fraction and the $C\!P$ asymmetry for $B^0 \to \omega K^{*}(892)^0$ result in
\be 
\tilde{f}^L_{\omega K^*} = 0.68\pm0.017\pm0.16, ~~~~ \mathcal{A}^L_{\omega K^*} = -0.13\pm0.27\pm0.13.
\ee

Triple Product Asymmetries (TPA) are also measured and are found to be below 5\%,  which is in agreement with theoretical predictions~\cite{predictionsForTPA}.

\section{Measurements of the $C\!P$ asymmetries in charmless four-body $\Lambda^0_b$ and $\Xi^0_b$ decays }

Despite theoretical predictions of about 20\% $C\!P$ violation for some charmless $\Lambda_b$ decays~\cite{predictionLb}, $C\!P$ violation was not observed in the baryon sector so far. The abundant production of $b$-hadrons at the LHC and the characteristics of the LHCb detector make this experiment particularly suitable to study the decays of these particles.

Six decay modes of $\Lambda_b^0, \Xi_b^0 \to p hh'h''$ are studied in this analysis, with 3fb$^{-1}$ of data collected by the LHCb experiment at centre-of-mass energies of 7~TeV and 8~TeV~\cite{baryons}. These decays proceed through $b \to u$ and $b \to s, d$ transitions. Experimental effects, such as detection and production asymmetries are cancelled out by computing the difference, $\Delta \mathcal{A}^{C\!P}$, between the raw values of the $C\!P$ asymmetries of the considered modes and the $C\!P$ asymmetries obtained in control modes, where no measurable $C\!P$ violation is expected. Further corrections are then applied to account for kinematical differences between signal and control modes and charge-dependent selection efficiencies. 

The results obtained for $\mathcal{A}^{C\!P}$ integrated over the whole phase space are:
\begin{equation*}
\resizebox{\linewidth}{!}{%
$\begin{aligned}
\Delta \mathcal{A}^{C\!P}(\Lambda_b^0 \to p \pi^{-} \pi^{+} \pi^{-}) & = (+1.1\pm2.5\pm0.6)\%,    ~~~
&\Delta \mathcal{A}^{C\!P}(\Lambda_b^0 \to p K{-} \pi^{+} \pi^{-}) & = (+3.2\pm1.1\pm0.6)\%,    \\
\Delta \mathcal{A}^{C\!P}(\Lambda_b^0 \to p K^{-} K^{+} \pi^{-}) & = (-6.9\pm4.9\pm0.8)\%,   ~~~
&\Delta \mathcal{A}^{C\!P}(\Lambda_b^0 \to p K^{-} K^{+} K^{-}) & = (+0.2\pm1.8\pm0.6)\%,   \\
\Delta \mathcal{A}^{C\!P}(\Xi_b^0 \to p K{-} \pi^{+} \pi^{-}) & = (-17\pm11\pm1)\%,  ~~~
&\Delta \mathcal{A}^{C\!P}(\Xi_b^0 \to p K{-} \pi^{+} K^{-}) & = (-6.8\pm8.0\pm0.8)\%.  \\
\end{aligned}$ }
\end{equation*}
%
In addition to the these inclusive results, measurements are also performed in specific regions of the phase space, for example at low two-body invariant mass or in quasi two- or three-body decay regions. A total of eighteen $C\!P$ asymmetries are measured and no significant $C\!P$ violation is observed in any of the measurements. 

A previous LHCb analysis~\cite{BaryonTPA}, performed with the same dataset, reported an evidence for $C\!P$ violation in a specific region of the phase space of $\Lambda_b^0 \to p \pi^{-} \pi^{+} \pi^{-}$ decay, using TPA while the present result shows no $C\!P$ violation for this mode. A comparison of the two results and methods can shed a light on the nature of this effect.

\section{Conclusion}

The LHCb collaboration has a very broad program of analyses searches for $C\!P$ asymmetries and the four analyses presented here only represent a small part of this program. During the presentation, measurements of the $C\!P$ asymmetry and branching fractions of $B^+ \to J/ \psi \rho^+$ obtained with run 1 data were also presented~\cite{Jpsi}: $\mathcal{B}(B^+\rightarrow J/\psi \rho^+)=\left( 3.81^{+0.25}_{-0.24} \pm 0.35 \right)\times 10^{-5}$ and $\mathcal{A}^{CP}(B^+\rightarrow J/\psi \rho^+)=-0.045^{+0.056}_{-0.057}\pm 0.008$. These results are the most precise ones form a single experiment to date. 

The run 2 data taking period is now over, and analyses with the full run 2 dataset, corresponding to an integrated luminosity of 6\invfb taken at a centre-of-mass energy of 13 TeV, are ongoing. The additional data sample will increase the sensitivity to the $C\!P$ observables and give access to more decay channels.





\end{document}